\documentclass[a4paper, UKenglish, cleveref, autoref, thm-restate]{oasics-v2021}

\usepackage{hyperref}
\usepackage{graphicx}
\usepackage{amsmath}
\usepackage{listings,multicol}
\usepackage{parcolumns}
\usepackage{lstcoq}
\usepackage{float}
\newcommand{\icode}[1]{\lstinline[mathescape,basicstyle=\small]!#1!}

\usepackage{xcolor}

\definecolor{ltblue}{rgb}{0,0.4,0.4}
\definecolor{dkblue}{rgb}{0,0.1,0.6}
\definecolor{dkgreen}{rgb}{0,0.5,0}
\definecolor{brightmaroon}{rgb}{0.76, 0.13, 0.28}
\definecolor{burntorange}{rgb}{0.8, 0.33, 0.0}
\definecolor{dkred}{rgb}{0.5,0,0}
\definecolor{bggray}{gray}{0.95}
\definecolor{arsenic}{rgb}{0.23, 0.27, 0.29}

\newcommand\coqref[2]{\raisebox{-0.3em}{\protect\coqLogo}{\footnotesize\href{https://github.com/AU-COBRA/ConCert/blob/#2}{\texttt{\color{brightmaroon}{#1}}}}}

\newcommand{\coqLogo}{\includegraphics[width=0.75em]{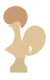}}

\title{Finding smart contract vulnerabilities with ConCert's property-based testing framework} 

\titlerunning{Finding smart contract vulnerabilities with ConCert} 

\author{Mikkel Milo}{Computer Science, Aarhus University}{mikkelmilo@gmail.com}{https://orcid.org/0000-0003-3261-5205}{}
\author{Eske Hoy Nielsen}{Computer Science, Aarhus University}{eske@cs.au.dk}{https://orcid.org/0000-0001-6735-6843}{}
\author{Danil Annenkov}{Computer Science, Aarhus University}{danil.v.annenkov@gmail.com}{https://orcid.org/0000-0001-8278-3069}{}
\author{Bas Spitters}{Computer Science, Aarhus University}{bas@cs.au.dk}{https://orcid.org/0000-0002-2802-0973}{}

\authorrunning{M. Milo, E. H. Nielsen, D. Annenkov and B. Spitters} 

\Copyright{Mikkel Milo, Eske Hoy Nielsen, Danil Annenkov and Bas Spitters} 

\supplementdetails[subcategory={Source Code}]{The ConCert Framework}{https://github.com/AU-COBRA/ConCert/tree/fmbc2022}

\nolinenumbers 
\hideOASIcs 

\EventEditors{Zaynah Dargaye and Clara Schneidewind}
\EventNoEds{2}
\EventLongTitle{4th International Workshop on Formal Methods for Blockchains (FMBC 2022)}
\EventShortTitle{FMBC 2022}
\EventAcronym{FMBC}
\EventYear{2022}
\EventDate{August 11, 2022}
\EventLocation{Haifa, Israel}
\EventLogo{}
\SeriesVolume{105}
\ArticleNo{1}

\begin{document}

\maketitle

\begin{abstract}
We provide three detailed case studies of vulnerabilities in smart contracts, and show how property based testing would have found them:
1. the Dexter1 token exchange;
2. the iToken;
3. the ICO of Brave's BAT token.
The last example is, in fact, new, and was missed in the auditing process.

We have implemented this testing in ConCert, a general executable model/specification of smart contract execution in the Coq proof assistant. 
ConCert contracts can be used to generate verified smart contracts in Tezos' LIGO and Concordium's rust language.
We thus show the effectiveness of combining formal verification and property-based testing of smart contracts.
\end{abstract}

\section{Introduction}
Blockchain-based technologies have seen rising interest in recent years.
This can be attributed to their ability to sustain a public distributed ledger with a high degree of reliability, integrity, and transparency, without requiring a trusted third party.
Smart contracts are distributed applications deployed on a blockchain.
They are typically used for sensitive transactions, for example, carrying large amounts of money or other valuable assets, but in principle, they can perform any computation.
Once a smart contract is deployed on the blockchain, it is impossible to change its source code.
The blockchain ensures that contracts are executed correctly according to the execution model.
However, it gives no guarantee that the smart contract's code is correct.
Like other programs, smart contracts are susceptible to bugs.

Some attacks on smart contracts have resulted in substantial losses.
For example, the ``DAO attack'' on Ethereum, where \$50 million worth of cryptocurrency was stolen due to a re-entrancy vulnerability\footnote{\url{https://www.wired.com/2016/06/50-million-hack-just-showed-dao-human/}}.
In April 2020, an attacker exploited a re-entrancy bug in the Lendf.me platform, resulting in a loss of about $99.5\%$ of the platform's funds ($\sim$\$25 million).
In 2021 cryptocurrency-related crimes including smart contract attacks resulted in losses of approximately \$14 billion~\cite{chainalysis}.
Hence, having a high assurance that a smart contract implementation is free of bugs is imperative.
Unit testing is often used in the process of smart contract development.
However, subtle bugs related to smart contract state evolution over a series of calls, or interaction with other contracts often cannot be captured by conventional unit testing.
Moreover, even proving functional correctness properties is not sufficient, as it was exemplified by the Dexter contract considered in~\cref{sec:dexter}.
To address such issues, we are using the ConCert framework in the Coq proof assistant which facilitates formal verification and property-based testing of smart contracts.
\paragraph*{Contributions.}

We present the details of the property-based testing functionality of the ConCert framework.
The testing functionality was presented briefly in earlier works on ConCert~\cite{ConCert:cpp2020,ConCert-extraction-testing:cpp2021}.
This paper contributes to the property-based testing functionality of ConCert and presents three case studies demonstrating how ConCert can be used to find real-world bugs in smart contracts.
Contributions to the testing framework include counterexample shrinking, negative testing capabilities, improved customisation and usability improvements.

The first two case studies show how ConCert could have been used to find bugs that were found in smart contracts by auditors and attackers.
The last case study shows how we used ConCert to find new bugs which could have led to upwards of \$8 million being stolen or frozen.

\section{ConCert Overview}
In this section, we give a brief overview of the ConCert framework, focusing on the smart contract execution layer and property-based testing.
ConCert is open-source, and available at \url{https://github.com/AU-COBRA/ConCert/}.

\subsection{Pipeline}
\begin{figure}
  \centering
  \includegraphics[width=12cm]{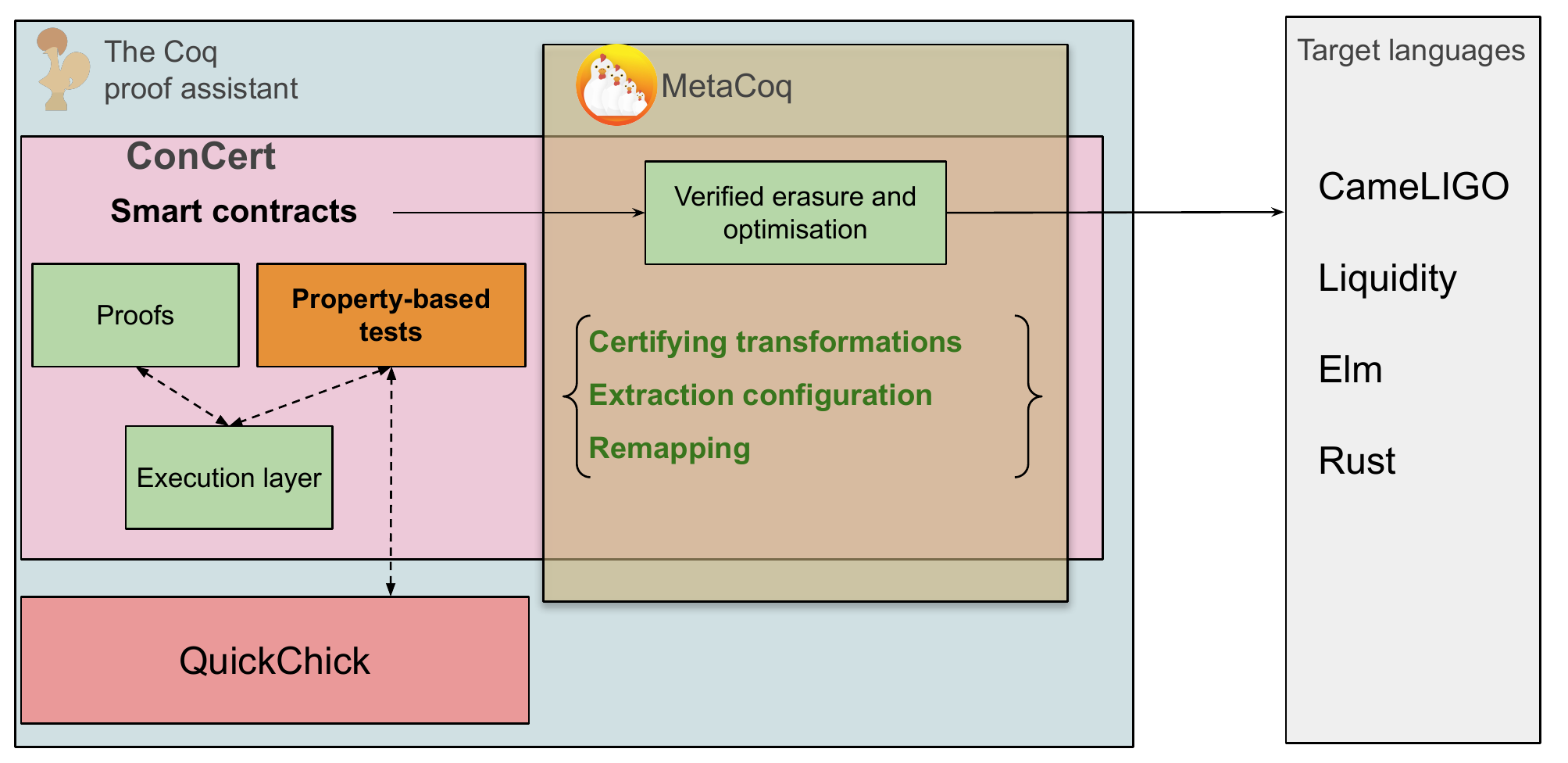}
  \caption{The pipeline}\label{fig:pipeline}
\end{figure}
\noindent%
The pipeline overview is presented in~\cref{fig:pipeline}.
We start by developing a smart contract in Coq using the ConCert infrastructure.
That is, smart contracts are written in Gallina, a functional language of Coq that shares similarities with other functional languages.
They are just ordinary functions that use some pre-defined blockchain primitives provided by the ConCert infrastructure.
This facilitates porting smart contracts written in functional smart contract languages to Coq.\footnote{E.g.\ LIGO, Liquidity, Sophia}
Even for a language like Solidity, this is fairly straightforward.
We then can write a specification and test the smart contact function semi-automatically against it, using the integration with QuickChick~\cite{denes2014quickchick}.
With more effort, we can also prove the properties of smart contracts using the ConCert infrastructure.
Proofs and tests crucially use the execution layer to reason about interacting contracts (see more details in~\cref{sec:exec-model}), which enables us to capture properties beyond the mere functional correctness of a single contract invocation (see~\cref{sec:dexter}).

After testing and verification, one can obtain an executable implementation in one of the supported smart contract languages through \emph{code extraction}.
Our development uses the verified erasure procedure of MetaCoq~\cite{MetaCoq} with verified optimisations and certifying pre-processing of ConCert.
This gives us a code-generation procedure with strong correctness guarantees and a small trusted computing base consisting of MetaCoq's \emph{quote} functionality, the pretty-printers into the target languages and the extraction configuration.
Note that ConCert's extraction does not use unsafe coercions, like \icode{Obj.magic} in OCaml.
Therefore, the resulting code is type-checked as a regular user-defined contract.
Additionally, extraction configuration involves mappings from ConCert's primitives to specific primitives for each supported target blockchain.
These mappings contribute to the TCB and are carefully defined together with experts for a particular target blockchain.
Outside of the ConCert pipeline, the compilers used to produce low-level code (e.g. Michelson) from extracted contracts are blockchain-specific and also contribute to the overall TCB.

\subsection{Smart Contract Execution Layer}\label{sec:exec-model}
The execution layer provides a model which facilitates reasoning about contract execution traces.
This makes it possible to state and prove temporal properties of interacting smart contracts.
Smart contracts in ConCert are modelled by abstracting a number of blockchains.\footnote{E.g. Concordium, Tezos, Dune, {\AE}ternity}
These blockchains can be characterised as variants of a message-passing model.
ConCert models core behaviour for such models.
Each blockchain can have some specific features not present in the ConCert execution model directly (e.g. Tezos' views), but similar behaviour can be expressed through message passing.
Contracts which use such features are not directly expressable in ConCert.
Some contracts might not be extractable to some targets if they use concepts that cannot be mapped to the target blockchain.

A contract consists of two functions:

\begin{itemize}
\item \icode{init : Chain -> ContractCallContext -> Setup -> option State}\\
The initialisation function is called after the contract is deployed on the blockchain.
The first parameter of type \icode{Chain} gives access to data about the blockchain (e.g.\ current chain height).
The \icode{ContractCallContext} parameter provides data about the current call (e.g.\ caller address, amount sent to the contract).
\icode{Setup} represents initialisation parameters.
\item \icode{receive : Chain -> ContractCallContext -> State -> option Msg ->}\\\icode{option (State * list ActionBody)}
  This function represents the main functionality of the contract that is executed for each call to the contract.
  \icode{Chain} and \icode{ContractCallContext} are the same as for \icode{init}.
  The parameter of type \icode{State} is the current state of the contract; \icode{Msg} is a user-defined type of messages that contract accepts (the \emph{entrypoints} of the contract).
  The result of a successful execution is a new state and a list of \emph{actions} represented with \icode{ActionBody}.
  The actions can be transfers, calls to other contracts (including itself), and contract deployments.
\end{itemize}
\noindent
Both \icode{receive} and \icode{init} are ordinary Coq functions, making them convenient to reason about.
However, reasoning about the contract functions in isolation is not sufficient, as many deployed contracts actually consist of a collection of interacting contracts, for example for the sake of modularity.
One call to \icode{receive} potentially emits more calls, which can create complex call graphs between deployed contracts.
Therefore, it is necessary to consider execution traces to prove some safety properties of smart contracts.
An execution trace \icode{ChainTrace} is the reflexive, transitive closure of a proof-relevant \icode{ChainStep} relation, which essentially captures the addition of a block to the blockchain.
In this step, any \emph{actions} (such as contract calls and transfers) coming from external users are executed.

\icode{ChainTrace} gives a relational operational semantics for the executions process.
The semantics is non-deterministic since it allows for arbitrary execution order for the actions emitted by contract calls.
Thus, ConCert provides two executable implementations: one follows depth-first and the other follows breadth-first order.
It also provides proof that if running \icode{add_block} succeeds, it results in a valid instance of \icode{ChainTrace}.
Having an executable implementation is crucial for property-based testing.

\subsection{Property-based Testing framework}\label{sec:pbt}
\begin{figure}
  \centering
  \includegraphics[width=8cm]{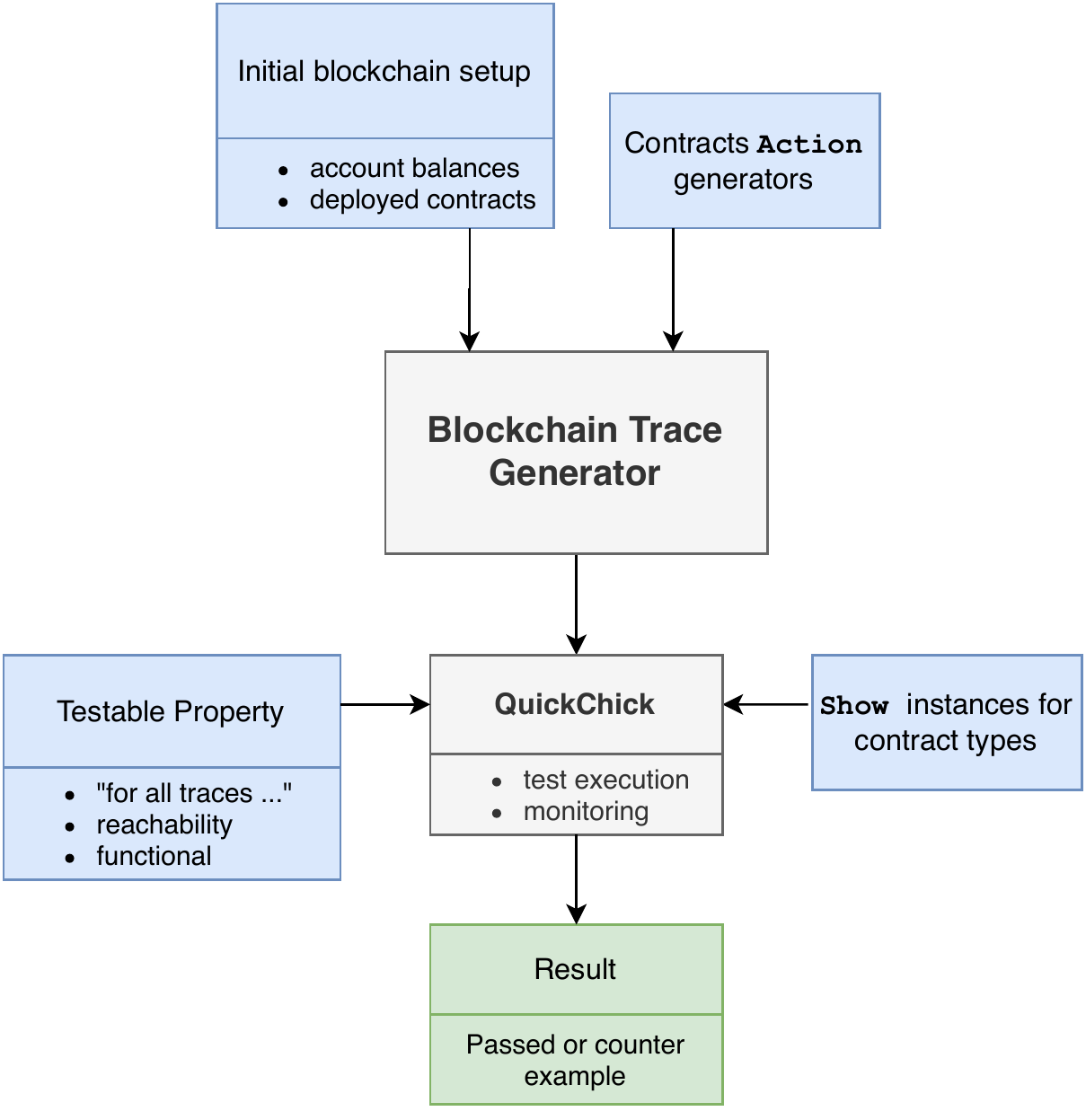}
  \caption{Property-based Testing in ConCert}\label{fig:testing-framework}
\end{figure}
\noindent%
Property-based testing (henceforth abbreviated \emph{PBT}), also known as \emph{random-property testing}, is a technique for testing where test data is generated pseudo-randomly and tested in large quantities against some decidable property.
We integrate the PBT library \emph{QuickChick}~\cite{denes2014quickchick} with the execution framework to obtain a method for testing contract executions.
In particular, we support testing the functional correctness of contracts but also testing (decidable) properties of entire execution traces.
The overview of the testing framework is given in~\cref{fig:testing-framework}.

In brief, the PBT framework works by having the user provide \emph{generators} for the \icode{Msg} type of the contract(s) tested.
In this context, generators are functions that produce pseudo-random values of the given type.
These generators are used to populate randomly generated execution traces with pseudo-random contract calls during testing with QuickChick.
The user also configures the initial blockchain setup consisting of account balances and contracts that are currently available for interaction (deployed contracts).
QuickChick also uses \icode{Show} type class instances to print test results (e.g.\ counterexamples).

For example, consider how to test a token contract whose \icode{Msg} type is
\begin{lstlisting}
Inductive Msg :=
    transfer of (address * address * nat)
  | approve of  (address * address * nat).
\end{lstlisting}
That is, it has two entrypoints: one for transferring tokens between the two given addresses and one for approving an address to spend a given number of tokens on behalf of another address.
Generating pseudo-random values of \icode{Msg} then amounts to either generating a \icode{transfer} or an \icode{approve}, and populating it with parameters by using the generators for \icode{address} and \icode{nat}.
We can either implement this manually or have QuickChick automatically derive such a generator\footnote{Due to limitations of QuickChick, the \icode{Derive} command fails for some parameterised inductive types, e.g. \icode{Msg} type in implicitly parameterised with some blockchain configuration. We have reported this issue: \url{https://github.com/QuickChick/QuickChick/issues/286}}.
Note that we might prefer to implement this manually since we might want to ensure that the number of tokens to be transferred in \icode{transfer} is never larger than the balance of the sender.
We provide various combinators to make it easy and convenient to implement complex generators.

Suppose we want to test that \icode{transfer} updates the internal balances correctly.
In ConCert, this functional correctness property is specified by using pre- and post-conditions.
Testing such a property with QuickChick could look like
\begin{lstlisting}
QuickChick ({{msg_is_transfer}} Token.receive {{transfer_correct}}).
\end{lstlisting}
The code above states that if the incoming message is a transfer, then after executing the token contract's \icode{receive} function, its state should be consistent with a predicate \icode{transfer\_correct}.
By default, QuickChick will generate 10.000 inputs and test that the property is satisfied in all of them, or otherwise report a counterexample.
The counterexamples reported are automatically minimized by the PBT framework to produce smaller counterexamples that are easy to understand.
From our experience, these tests typically take less than a minute (see~\cref{sec:evaluation}).

One can also test whether some state is reachable from the given state.
For example, the following test
\begin{lstlisting}
QuickChick (token_cb $\sim\sim$> (person_has_tokens person_3 42)).
\end{lstlisting}
shows that from the state \icode{token_cb} with three addresses participating in the token there is a state where \icode{person_3} has 42 tokens.
The corresponding trace is reported to the user.

\section{Dexter decentralized exchange}\label{sec:dexter}
In this section, we consider a bug in (an earlier version of) Dexter, a decentralized token exchange contract on the Tezos blockchain.
The bug would have allowed an attacker to manipulate exchange rates to obtain unintended profit through a simple attack.
The contract had previously been formally verified for functional correctness\footnote{\url{https://research-development.nomadic-labs.com/dexter-decentralized-exchange-for-tezos-formal-verification-work-by-nomadic-labs.html}}.
However, this bug can only be discovered when considering \textit{execution traces} - that is, sequences of contract calls.
We demonstrate how this bug can be found by testing a \emph{natural} specification on traces.
So, we argue that this bug would likely have been discovered when using ConCert as part of the specification process.

The Dexter exchange smart contract is used for exchanging tokens and tez (the on-chain currency of Tezos), it implements a so-called \textit{constant-product market}, which means that the total value of the contract never decreases.
A property of such markets is that the exchange rate cannot be significantly manipulated unless a party owns most of the market's assets~\cite{angeris2019analysis}.
The rate at which tokens and tez can be exchanged is calculated dynamically at each trade according to the function
\begin{align*}
  getInputPrice(Ts, Ts_{reserve}, Tez_{reserve}) = \frac{Ts \cdot 997 \cdot Tez_{reserve}}{Ts_{reserve} \cdot 1000 + Ts \cdot 997}
\end{align*}
where $Ts$ are the tokens being exchanged, $Ts_{reserve}$ is the reserve of tokens held by the Dexter contract, and $Tez_{reserve}$ is the contracts tez reserve.

One key property of constant-product markets, that cannot be verified from functional correctness alone, is that splitting trades is never profitable.
Specifically, suppose a user trades $N$ tokens for $Z$ tez.
Suppose this trade is split into $k > 1$ trades, totalling $N$ tokens for a total of $Z'$ tez.
Then it should be the case that $Z' \le Z$.

In ConCert, we can state this property by asserting that for each block added to generated traces, the total amount of tez gained from trades does not exceed what the user would have gained from trading the same amount of tokens in a single exchange. The full Coq definition can be found in \coqref{examples/dexter/DexterTests.v}{fmbc2022/examples/dexter/DexterTests.v\#L97}

With this test, our PBT framework automatically finds a counterexample that violates the property.
The counterexample show two consecutive exchanges; first trading 14 tokens for 5 tez, then 16 tokens for an additional 5 tez.
However, the payout for a single trade of 30 tokens would have been 9 tez, netting the user an extra one tez from splitting the trade.
The vulnerability is due to a combination of Tezos' breadth-first execution model\footnote{Tezos moved to depth-first execution order after Dexter was developed} and the way the contract tracks its asset reserves.
Concretely the problem is that in breadth-first both trades are executed before the actions emitted by the trades are executed, meaning that the second trade will start before the tez and tokens from the first trade have finished being transferred.
The contract accounts for this by manually tracking the number of tokens, but fails to do the same for the tez reserve.
Thus when the second trade starts the contract uses the wrong tez reserve for calculating the exchange rate.
A strength of ConCert is that it allows testing in both depth-first and breadth-first execution order, running the same test with depth-first shows no vulnerability.

The bug was fixed prior to the deployment of Dexter.

\section{iToken}\label{sec:iToken}
In this section, we show how the bZx iToken smart contract was compromised and how ConCert could have discovered this vulnerability.
The iToken smart contract is an interest accumulating ERC20 token used as part of the bZx decentralized finance platform.
In September 2020 an attacker stole \$8 million worth of cryptocurrency by exploiting a vulnerability in the iToken contract caused by a misplaced line of code\footnote{\url{https://bzx.network/blog/incident}}.
This vulnerability was missed by two audits of the platform.
The vulnerability was in the tokens \icode{transferFrom}, which is used to transfer tokens between users.
The transfer logic was implemented in the following way:
\begin{lstlisting}
uint256 balanceFrom = balances[from];
uint256 balanceTo   = balances[to];
balances[from] = balanceFrom.sub(amount);
balances[to]   = balanceTo.add(amount);
\end{lstlisting}
This logic would have been safe had lines 2 and 3 been swapped.
To see where this goes wrong, consider the case where \icode{from = to}.
In this case, the transferred amount would be subtracted from the sender's balance in line 3.
However, in line 4 the original balance of the sender is used to add the transferred amount to the sender's balance, resulting in the sender ending gaining tokens through the self-transfer.

This bug could be found using the PBT framework by writing a test checking that the balance remains the same after a self-transfer.
However, such a test would require knowledge of the possibility of a bug in this edge case.
Instead, we formulate the property that \emph{the sum of all balances should remain unchanged after a call}, with the exception of minting and burning calls.
In ConCert testing such a property looks like:
\begin{lstlisting}[language=Coq,title={\coqref{examples/iTokenBuggy/iTokenBuggyTests.v:sum\_balances\_unchanged}{fmbc2022/examples/iTokenBuggy/iTokenBuggyTests.v\#L98}}]
Definition msg_is_not_mint_or_burn state msg :=
  match msg with
  | mint _ | burn _ => false
  | _ => true
  end.
Definition sum_balances_unchanged chain cctx (old_state : State) (msg : Msg)
                                          (result : option (State * list ActionBody)) : bool :=
  let balances_sum state := sum s.(balances) in
  match result with
  | Some (new_state, _) => balances_sum old_state =? balances_sum new_state 
  | None => true (* Return true in the case that nothing changed *)
  end.

QuickChick ({{msg_is_not_mint_or_burn}} iTokenContract {{sum_balances_unchanged}})
\end{lstlisting}
By running the test, we indeed obtain a minimal counterexample showing that self-transfers violate the property.

\section{Basic Attention Token}\label{sec:BAT}
In this section, we show how ConCert was used to find new bugs, that were missed by several audits, in the Basic Attention Token (BAT) smart contract.
BAT is an Ethereum initial coin offering smart contract developed by Brave.
It is a combination of an ERC-20 token and a crowdsale contract, where users can fund ether to Braves' project in return for BAT tokens.
The crowdsale runs for a fixed amount of blocks, after which the funding either succeeds or fails.
If funding succeeds, Brave receives all the ether raised.
If it fails, all users can claim a refund of their ether by burning their tokens.
As the contract owners, Brave get a fixed amount of free tokens to spend.

We test functional correctness using a similar Hoare triple test as shown in ~\cref*{sec:pbt}.
In addition, we formulated five key safety properties.
\begin{enumerate}
  \item \textbf{Funding is final:} Once the contract enters its funded state it cannot leave it again.
  \item \textbf{Funding possible:} If there is enough ETH in the blockchain to reach the funding goal, then it should be possible to reach a state in which the funding succeded.
  \item \textbf{No refunding for owners:} The free tokens given to the owners should not be refundable.
  \item \textbf{Refund guarantee:} There should always be enough ETH in the contract balance to refund all funded tokens.
  Unless funding succeded.
  \item \textbf{No frozen funds:} It should always be possible to completely drain the contract balance, so no ETH gets permanently frozen.
\end{enumerate}
Through testing, we found that only the first property holds.
Most of the bugs occur from combining token and crowdsale functionality and both parts behave safely on their own.
\emph{This highlights that composing contracts is nontrivial and can easily introduce subtle bugs.}

\subsection{Test Setup}
In~\cref{sec:dexter,sec:iToken} we showed that ConCert could find known bugs.
For those, it was not so important whether the generators would cover the entire input space.
However, when testing a complex contract with the purpose of finding potentially unknown bugs, it is crucial to have good generators.
A good quality generator should be able to cover the entire input space of the smart contract and have a good balance between generating calls that succeed and calls that fail.
Using automatically derived generators will often result in too many failing calls for complex smart contracts.
For testing BAT we take the approach of combining manually written generators designed to only produce valid calls with generators that are likely to produce invalid calls.
That is, for each entrypoint, we define two generators.
This is illustrated in~\cref{fig:gFinalize}.
The \icode{finalize} entrypoint is an entrypoint that transitions the contract from funding to the funded state.
It can only be called by the owner after funding succeeds.
The first generator \icode{gFinalize} only produces calls that we expect to succeed, while the \icode{gFinalizeinvalid} generator will generate calls with an arbitrary sender, which is unlikely to be valid.
We use the \icode{x <- e1 ;; e2} monadic bind notation to bind generated values.
The generators for potentially invalid calls can be automatically derived using QuickChick.
All the generators are combined into a single call generator.

This approach gives us a generator that can cover the entire input space while still allowing us to tune the distribution of valid and invalid calls to different entrypoints.
Using the PBT framework we can measure statistics about the generator and use that to tune the distribution.

\begin{figure}
\begin{lstlisting}[language=Coq,title={\coqref{examples/bat/BATGens.v:gFinalize}{fmbc2022/examples/bat/BATGens.v\#L82}}]
Definition gFinalize env contract_state : G (option (Address * Msg)) :=
  if (isFullyFunded env contract_state) (* Check if funding succeded *)
  then returnGen (Some (fund_addr, finalize)) (* Call finalize from owner address *)
  else returnGen None. (* Don't return call if not funded *)
Definition gFinalizeInvalid env contract_state : G (Address * Msg) :=
  sender <- gAddress ;; (* Generate arbitrary address *)
  returnGen (sender, finalize).
\end{lstlisting}
  \caption{Generators for the \icode{finalize} entrypoint}\label{fig:gFinalize}
\end{figure}

\subsection{Finding Vulnerabilities}
We test each of the five safety properties for the BAT contract defined in~\cref{sec:BAT}.
Here we detail a few of the tests.

A key property is that the contract doesn't deadlock, i.e. with enough user support it should always be possible to reach the funded state.
Since ConCert can test reachability of states we can easily state this property by combining the reachability checker with a deployment configuration generator.
The following test states that for any BAT deployment configuration there should exist a trace from the state where BAT is deployed with that configuration to a state where the contract is funded.
\begin{lstlisting}[language=Coq,title={\coqref{examples/bat/BATTests.v}{fmbc2022/examples/bat/BATTests.v\#L1118}}]
QuickChick (forAll gBATSetup (build_init_cb (fun cb => cb ~~> is_finalized))).
\end{lstlisting}
Here \icode{gBATSetup} is the configuration generator, \icode{build_init_cb} builds an inital state with the contract deployed, and \icode{is_finalized} checks for a given blockchain state if the contract is funded.
By running the test, we obtain counterexamples showing four classes of configurations where the contract cannot be fully funded.
One of them is the case where the funding period is empty or already over at the time of deployment.
Ideally, the contract should have included a check at deployment preventing such configurations.

A crucial safety property is that any user who donated should be guaranteed their money back in case of failed funding.
By testing the functional correctness of entrypoints, we already know that the contract will always refund the correct amount and will always succeed, given that the contract has enough funds.
Therefore, testing refund guarantee reduces to checking that there is always enough funds to refund all tokens held by "real" users.
Here we distinguish between real users of the contract and the owner, because the owner's free tokens should not be counted.
That is, we want to test that the following is always true.
$$\texttt{contractBalance}\ge\frac{totalTokenSupply - ownersTokens}{tokenExchangeRate}$$
In ConCert a test of this looks like:
\begin{lstlisting}[language=Coq,title={\coqref{examples/bat/BATTests.v:contract\_balance\_lower\_bound}{fmbc2022/examples/bat/BATTests.v\#L801}}]
Definition contract_balance_lower_bound (cs : ChainState) :=
  let contract_balance := env_account_balances cs contract_base_addr in
  (* Get BAT contract state *)
  match get_contract_state State cs contract_base_addr with
  | Some cstate =>
    (* Get token balance of owner *)
    let bat_fund_balance := with_default 0 (FMap.find owner (balances cstate)) in
      if cstate.(isFinalized)
      then checker true (* Case where refunds are not permitted *)
      (* Assert that there is enough ETH to refund all tokens held by "real" users *)
      else checker (Z.geb contract_balance
        (Z.of_N (((total_supply cstate) - bat_fund_balance) / cstate.(tokenExchangeRate))))
  | None => checker true (* Case where contract isn't deployed *)
  end.
QuickChick (forAllChainState contract_balance_lower_bound)
\end{lstlisting}
Running the test we get the following minimized counterexample from the testing framework.
\begin{lstlisting}
Chain{|
  Block 1 [Action{act_from: 10, act_body: (act_deploy 0, Setup{owner:=17;...})}];
  Block 2 [Action{act_from: 17, act_body: (act_call 128, 0, transfer 16 14)}]
|}
\end{lstlisting}
This counterexample shows a trace where the BAT contract is deployed in the first block, after which the owner (address 17) immediately transfers some of its free tokens to another user.
This is possible because the contract combines crowdsale and token contract behaviour.
This violates two of the safety properties because nothing is preventing the second user from refunding the transferred tokens.
Thus it is possible for the free tokens given to the owner to be refunded by first transferring them.
This also breaks the property that all real users should be guaranteed a refund because if the owner refunds some of the free tokens then there is no longer enough ETH to refund all tokens held by real users.

The remaining safety properties were tested using similar methods.

\section{Related Work}

Various testing approaches have been applied to smart contracts.
Tools like Truffle\footnote{\url{https://trufflesuite.com/}} for Ethereum or SmartPy\footnote{\url{https://smartpy.io/docs/scenarios/testing/}} for Tezos mostly cover conventional unit testing that can be insufficient.
The testing framework for LIGO\footnote{\url{https://ligolang.org/docs/advanced/testing}} supports unit testing and mutation testing.
However, none of the conventional testing frameworks offers a possibility for generating random traces and testing properties of interacting contracts.
We will now focus on works using fuzzing/property-based testing techniques.

The closest to our work is the property-based testing framework for the Tezos' Michelson language.
The framework utilises QCheck, a QuickCheck-inspired property-based testing framework for OCaml.
QCheck was extended by Nomadic Labs with the ability to generate arbitrary sequences of Liquidity Baking contract calls.
The contract is manually reimplemented in OCaml and serves as a model for the original contract.
The model implementation is then validated against the original contract through the actual Tezos execution model.
The development is tailored to the Liquidity Baking contract and is not connected to the Michelson formalisation in Coq Mi-Cho-Coq~\cite{Mi-Cho-Coq}.
We are currently collaborating with the Mi-Cho-Coq team on integrating ConCert with the formalisation of Michelson.

For the Ethereum blockchain, several works are using randomised testing techniques.
Echidna~\cite{grieco:Echidna} and Brownie\footnote{Property-based testing framework for EVM: \url{https://github.com/eth-brownie/brownie}} use fuzzing-like techniques for testing smart contracts.
The common challenge for this approach is that randomly generated transaction data might not be enough to ensure good coverage.
This is especially problematic in the case of smart contract interactions, since the whole sequence (trace) of actions must be generated.
Echidna uses coverage-driven feedback to automatically tune the testing parameters.
Brownie uses unit-test like tests with user-defined generators for randomising inputs to contract calls in the tests.
Brownie does not generate calls or execution traces, which limits the types of bugs that it can find.
In our approach, instead of tuning pre-defined parameters, we allow users to define generators that produce random data with fewer discarded tests.
For simple cases, data generators can be derived automatically using the QuickChick infrastructure.

The EthPloit project~\cite{Zhang:EthPloit} generates possible exploits using fuzzing techniques.
The exploits are split into three categories.
For each of these categories, a special exploit detector oracle is used to report an exploit.
For example, the Balance Increment oracle compares the overall initial balance of attackers' accounts with the current balance after a series of transfers and reports, if the balance of the attackers' accounts increases.
EthPloit utilises static analysis to focus attention on particular variables and functions.
The input for selected functions is generated randomly, or chosen using a seed set.
The seed sets are used to provide runtime feedback.
This improves the fuzzing efficiency by exploiting the results of previous runs.
In our approach, the users specify the properties to test, instead of searching for particular categories of exploits.
Violation of such properties is reported as a counterexample, which points to vulnerabilities.
The pure/functional nature of our smart contracts avoids many pitfalls and simplifies reasoning about smart contracts.
When compared to effectfull languages, such as Solidity, static analysis is less urgent.

Finally, the \texttt{cooked-validators} library\footnote{\url{https://iohk.io/en/blog/posts/2022/01/27/simple-property-based-tests-for-plutus-validators/}} for the Plutus smart contract language~\cite{Chapman:PlutusCore} facilitates property-based testing with arbitrary transaction sequences.
Note, however, that the execution model for Plutus does not involve on-chain inter-contract communication.
Plutus itself supports property-based testing at the contract endpoint level using QuickCheck.\footnote{\url{https://plutus-pioneer-program.readthedocs.io/en/latest/pioneer/week8.html\#using-quickcheck-with-plutus}}

\section{Evaluation}\label{sec:evaluation}
We evaluate our framework in terms of usability, specifically regarding bug-finding capabilities.
We demonstrated the testing framework on three concrete examples in the previous section, showing that it can find different types of real-world bugs.
The vulnerabilities had a wide range of causes: the execution order, complex contract-to-contract interactions and the evolution of the contract state.
Such bugs would not have been detected in other tools considering only functional correctness.
This highlights ConCert's unique capability of modelling and testing complex contract interactions.

We have tested various other smart contracts, such as a reference implementation of the ERC-20 Token\footnote{\url{https://github.com/AU-COBRA/ConCert/tree/master/examples/eip20}}, and re-discovered known bugs, thus supporting the claim that our framework is effective at finding bugs.
Since we have the full power of Coq at our disposal, we can effectively test any \textit{decidable} property on the \icode{Chain} type.
Hence, there are few limitations in terms of expressiveness.
While ConCert can find many common bug types, some bugs, such as vulnerabilities related to gas, remain out of reach of ConCert.

We also emphasise that once contracts are implemented (in ConCert) and the executable specifications are written (i.e.\ the decidable properties to be proven or tested), the only prerequisite for automatically testing the specifications is to implement the action generators and show instances, as discussed in~\cref{sec:pbt}.
Implementing these requires only some expertise with Gallina and QuickChick, and can in some cases be derived automatically.
Hence, the setup is relatively simple, only requiring moderate extra effort compared to writing traditional tests for users already familiar with property-based testing and Gallina or similar functional languages.

Since the contracts tested were ported to ConCert there is the risk that bugs were introduced in this process.
However, since the framework gives full counterexamples it is easy to verify that bugs found are also present in the original contract, this part could also be automated.
Another worry could be that the implementations of the contracts or the generators were tailored to finding known bugs.
We took extra care implementing the generators for all three contracts, making sure that no knowledge of known bugs was used.
Moreover, we did not test for a specific bug but for natural properties that would be part of any reasonable specification.
For the Dexter and iToken contracts, we only implemented the entrypoints related to the known bugs.
This slightly sped up finding the bugs, but adding the other entrypoints would only slow this down by a small constant factor.
For the BAT contract the full contract was ported and it was not tailored towards any specific properties.

Additionally, the feedback loop from executing tests is fast, making it easy to use during the contract development process.
In our experience, QuickChick will usually report counterexamples, if they exist, within 1-2 seconds and otherwise report that all inputs (by default 10.000 traces) passed --- usually in 5-10 seconds (for traces of 14 calls).
Of course, the time depends on many factors, most importantly, the length of traces and the complexity of generators and contracts.
Heuristically, we limit ourselves to 10.000 tests, based on the extensive experience from QuickChick.
Naturally, tests cannot fully guarantee that there is no bug, we use proofs for that goal.

\section{Conclusions}
We have presented the ConCert Coq framework for testing, verifying and extracting smart contracts.
We have demonstrated the framework for property-based testing on three smart contracts using it to discover vulnerabilities used in previous attacks and new bugs that could have led to millions of dollars stolen or frozen.
As stated in the previous section, the vulnerabilities had a wide range of causes covering the most common causes of flaws in smart contracts.

We have re-discovered several bugs in real-world contracts (not presented in this paper), such as the \$50 million ``DAO attack'' on Ethereum, and tested reference implementations of ERC-20 and FA2 Token Standards, common standards for tokens used in several blockchains\footnote{\url{https://github.com/AU-COBRA/ConCert}}.

Hence, our approach to testing smart contracts scales to real-world contracts and is capable of finding significant bugs.
Contracts in ConCert are extractable to Concordium's Rust framework, Liquidity, and CameLIGO.
Thus in total, we have a toolchain for producing executable code for smart contracts that are tested and verified.
The importance of combined auditing, testing and verification is also starting to be recognized by the industry.\footnote{e.g.\ \url{https://forum.cardano.org/t/cip-proposal-cardano-audit-best-practice-guidelines/100022}}

\section*{Acknowledgements}%
We would like to thank the LIGO team and in particular Tom Jack, Rapha\"el Cauderlier, Exequiel Rivas, R\'emi Les\'en\'echal, Gabriel Alfour, Thomas Letan and Arvid Jakobsson for the discussions about testing for LIGO.
This research was partially supported by a grant from Nomadic Labs and by the Concordium Blockchain Research Center.



\bibliography{bibliography}

\end{document}